\documentclass[journal]{IEEEtran}
\usepackage{xcolor,soul,framed} 

\colorlet{shadecolor}{yellow}
\usepackage[pdftex]{graphicx}
\DeclareGraphicsExtensions{.pdf,.jpeg,.png}

\usepackage[cmex10]{amsmath}
\usepackage{array}
\usepackage{float}
\usepackage{subcaption}  
\usepackage{mdwmath}
\usepackage{mdwtab}
\usepackage{eqparbox}
\usepackage{url}
\usepackage{tikz}
\usetikzlibrary{arrows.meta}
\usepackage{multirow}
\usepackage{array}
\usepackage{longtable}
\usepackage{flushend}

\begin{document}

\bstctlcite{IEEEexample:BSTcontrol}
    \title{Enhancing Autonomous Driving Safety Analysis with Generative AI: A Comparative Study on Automated Hazard and Risk Assessment}
  \author{Alireza Abbaspour,~\IEEEmembership{ Senior Member,~IEEE,} Aliasghar Arab$^*$,~\IEEEmembership{Senior Member,~IEEE,} Yashar Mousavi,~\IEEEmembership{ Member,~IEEE.}
\thanks{A. Abbaspour is Staff System Safety Engineer at Qualcomm (email: aabba014@fiu.edu), A. Arab is with the Department of Mechanical and Aerospace Engineering at New York University, Brooklyn, NY, USA and Department of Electrical Engineering at North Carolina A\&T State University, Greensboro, NC, USA, (email: aliasghar.arab@nyu.edu; aarab@ncat.edu), Y. Mousavi is an Alumni from Department of Applied Science, School of Computing, Engineering and Built Environment, Glasgow Caledonian University, Glasgow, UK and currently works at American Axle \& Manufacturing (email: seyedyashar.mousavi@gcu.ac.uk). }
}

\maketitle
\begin{abstract}
The advent of autonomous driving technology has accentuated the need for comprehensive hazard analysis and risk assessment (HARA) to ensure the safety and reliability of vehicular systems. Traditional HARA processes, while meticulous, are inherently time-consuming and subject to human error, necessitating a transformative approach to fortify safety engineering. This paper presents an integrative application of generative artificial intelligence (AI) as a means to enhance HARA in autonomous driving safety analysis. Generative AI, renowned for its predictive modeling and data generation capabilities, is leveraged to automate the labor-intensive elements of HARA, thus expediting the process and augmenting the thoroughness of the safety analyses.
Through empirical research, the study contrasts conventional HARA practices conducted by safety experts with those supplemented by generative AI tools. The benchmark comparisons focus on critical metrics such as analysis time, error rates, and scope of risk identification. By employing generative AI, the research demonstrates a significant upturn in efficiency, evidenced by reduced timeframes and expanded analytical coverage. The AI-augmented processes also deliver enhanced brainstorming support, stimulating creative problem-solving and identifying previously unrecognized risk factors.
The findings highlight the potential of generative AI to transform safety engineering paradigms by streamlining workflows, mitigating oversight, and broadening the horizon of hazard perception. The paper argues for integrating generative AI in the early stages of safety system development to identify and address latent system vulnerabilities preemptively. In doing so, it posits a proactive safety culture that aligns with the dynamic complexities of autonomous driving technologies. The paper concludes by discussing the implications for future safety engineering practices and the pivotal role of generative AI in establishing rigorous, fault-tolerant autonomous driving systems.
\end{abstract}

\section{Introduction}
As we advance into the future with autonomous driving technologies, the role of meticulous safety analysis cannot be overstated. Autonomous vehicles, with their sophisticated algorithms for navigation, decision-making, and interaction with the environment, introduce a complex layer of potential risks and uncertainties. Conducting thorough safety analyses is essential for identifying, understanding, and mitigating these risks, ensuring that autonomous driving systems can be safely integrated into our society. This rigorous approach to safety is not only a technical requirement but a foundational element in building public trust and acceptance \cite{koopman2022ul}. As autonomous driving systems move from research laboratories to public roads, the importance of safety analysis escalates, becoming a critical gateway to commercial success. Without demonstrating a robust commitment to safety through comprehensive analysis, the path to commercializing autonomous vehicles faces significant hurdles, from regulatory approvals to consumer adoption. Thus, safety analysis stands at the core of the journey toward a future dominated by autonomous driving, underpinning the technology's viability, reliability, and acceptance in society.

\noindent Within the safety analysis framework for Autonomous Driving Systems (ADS), HARA plays a pivotal role \cite{beckers2013structured}. It serves as a systematic process to identify potential hazards, assess the associated risks, and implement safety concepts to mitigate these risks effectively. HARA is instrumental in ensuring that every conceivable scenario where the ADS could fail or contribute to unsafe conditions is carefully analyzed. This proactive approach allows for the design of robust safety measures, such as fail-safe mechanisms and redundancy, which are integral to enhancing the overall safety of the system.

\noindent However, conducting HARA in the context of ADS has its own challenges \cite{beckers2017structured, khastgir2017towards}. The complexity and diversity of potential driving scenarios mean that the process can be incredibly time-consuming and labor-intensive. Moreover, the reliance on human analysts introduces the possibility of human error, where subjective judgments or oversights could lead to underestimating certain risks or completely missing others. The voluminous nature of the task, coupled with the rapid pace of technological advancements in autonomous driving, further intensifies these challenges. Addressing these issues demands not only a meticulous approach to HARA but also the exploration of advanced tools and methodologies that can streamline the process, reduce the potential for human error, and manage the vast workload effectively, thereby ensuring a thorough and reliable safety analysis.

\noindent These challenges motivated us to propose a solution for the automation of HARA and reduce the required workload for this process while improving its accuracy. The recent advancement in Generative Artificial Intelligence (Gen-AI) was a key enabler for us in the automation of HARA.
Generative AI represents a groundbreaking shift in the landscape of technology and its applications across various domains of knowledge. The development of Generative AI began as an exploration into creating models that can generate new data points, mimicking the distribution of real data. Over time, this exploration has culminated in advanced algorithms capable of producing text, images, music, and even synthetic data that closely resemble their real-world counterparts. The applications of Generative AI are vast, ranging from content creation in the digital media industry to drug discovery in pharmaceuticals, and from predictive maintenance in manufacturing to personalized learning in education. The core appeal of Generative AI lies in its ability to automate complex tasks, enhancing efficiency, creativity, and decision-making processes. Hao et al. developed a natural adversarial scenario generation platform to create large-scale, diverse, and realistic safety-critical test scenarios for AV testing \cite{hao2023adversarial}. They integrated rule-based models with generative adversarial imitation learning to improve the efficiency of generating critical safety scenarios. In another study \cite{krajewski2018data}, the authors utilized an unsupervised Info-GAN architecture to develop models capable of generating realistic vehicle trajectories for ADS safety validation. 

\noindent By training the developed generative AI-based model on an extensive amount of measured lane change trajectories from the dataset, the model was reported able to generate new, realistic lane change maneuvers based on a small set of intuitive parameters. An edge intelligence-enabled generative AI framework for vehicular networks was developed to address critical vehicular challenges by proposing a workflow for collaborative fine-tuning and distributed inference \cite{xie2024gai}. As reported, the developed framework improved content generation by leveraging the collaboration between roadside units and vehicles, thus improving efficiency and capability in dynamic vehicular scenarios. A generative world model for autonomy was introduced by Hu et al. \cite{hu2023gaia}, capable of creating realistic driving scenarios from video, text, and action inputs, offering detailed control over vehicle behavior and scene features. The investigated approach treated world modeling as an unsupervised sequence modeling task, enabling the model to learn scene dynamics and spatial relationships, thereby advancing autonomous driving training. Xu et al. \cite{xu2023generative} developed an autonomous driving framework to enhance driving safety and traffic management efficiency. 

\noindent A generative AI-based platform was employed to create extensive simulated traffic and driving data for offline training purposes. In another study \cite{lateef2021saliency}, a framework employing generative adversarial networks (GAN) was developed to predict significant objects in driving scenes, thereby improving technologies such as ADAS and autonomous driving systems. The authors built a large-scale visual attention-driving database of saliency heat-maps from existing driving datasets to effectively predict the most significant objects in a driving environment.

\noindent The automation of HARA for ADS is a prime example of the potential of Generative AI to revolutionize traditional processes.  Generative AI, with its capacity to analyze vast datasets and simulate a multitude of driving scenarios, offers a compelling solution to these challenges. By automating the HARA process, Generative AI can significantly reduce the time required to conduct safety analysis, while also improving its accuracy and comprehensiveness.

\noindent Recently, researchers have been investigating approaches to integrate generative AI models, e.g., ChatGPT to different safety analysis algorithms \cite{uddin2023leveraging,diemert2023can, qi2023safety}. In \cite{uddin2023leveraging}, they investigated whether ChatGPT could improve hazard recognition among construction students. The study conducted with 42 students at a major U.S. university found that integrating ChatGPT into the curriculum significantly enhanced students' hazard recognition abilities, suggesting its potential benefits in safety education.

\noindent The strength of Generative AI in automating HARA lies in its ability to generate and evaluate countless scenarios that might be too complex, rare, or simply overlooked by human analysts. This includes the simulation of extreme conditions, unpredictable human behavior, and novel interactions between vehicles and their environment. Furthermore, Generative AI can continually learn and adapt to new data, ensuring that the safety analysis evolves in alignment with advancements in ADS technologies and changes in driving environments. By leveraging Generative AI, the HARA process becomes not only faster but also more robust, enabling a more precise identification of risks and the development of effective mitigation strategies. Thus, Generative AI stands as a powerful tool to enhance the safety analysis of autonomous driving systems, marking a significant step forward in the journey towards safer and more reliable ADS technologies.

\noindent In the following sections, the proposed solution of automating HARA using generative AI is explained, and a comprehensive study is presented to show the ability of automated HARA to do safety analysis of ADS. Section II explains the overall HARA process and defines the main steps for conducting HARA. Section III illustrates our proposed solution for automating HARA. Section IV presents a case study that we performed to show the automation process. 

\section{HARA Standard Framework}
\label{sec:2}

HARA has a rich history rooted in various industries, where its principles have been refined and adapted to address the evolving challenges of safety management. The origins of modern HARA can be traced back to the fields of aerospace and nuclear engineering in the mid-20th century, where the need to mitigate complex risks associated with highly technical systems became increasingly apparent \cite{covello1985risk}.

\noindent In the aerospace industry, HARA gained prominence during the development of manned spaceflight programs, such as NASA's Mercury, Gemini, and Apollo missions. Engineers and safety experts recognized the need for systematic methods to identify and analyze potential hazards in spacecraft design and operation. This led to the development of techniques such as Failure Modes and Effects Analysis (FMEA) and Fault Tree Analysis (FTA), which became foundational tools in the field of hazard analysis. Over time, the principles and techniques of HARA have been adapted and applied to a wide range of industries, including automotive, pharmaceuticals, healthcare, and information technology. 

\noindent Today, as autonomous driving continues to advance, HARA remains at the forefront of efforts to ensure the safety and reliability of these systems. By drawing on decades of experience and expertise from various industries, engineers, and researchers are continually refining HARA methodologies to address the unique challenges posed by autonomous vehicles and pave the way for their safe and widespread deployment.

\noindent ISO 26262, an international standard for automotive functional safety, outlines a systematic approach to managing safety throughout the automotive development lifecycle. Central to this approach is the concept of HARA, which is a key requirement of the standard.
HARA, conducted early in the development process, identifies potential hazards and assesses associated risks in vehicle functions, interfaces, and operating conditions. By adhering to ISO 26262's HARA recommendations, automotive manufacturers systematically mitigate safety risks, enhancing vehicle safety and reliability to achieve an acceptable risk level, quantified through the automotive safety integrity level (ASIL). 

\begin{figure*}
    \centering
 \includegraphics[width=0.9\textwidth]{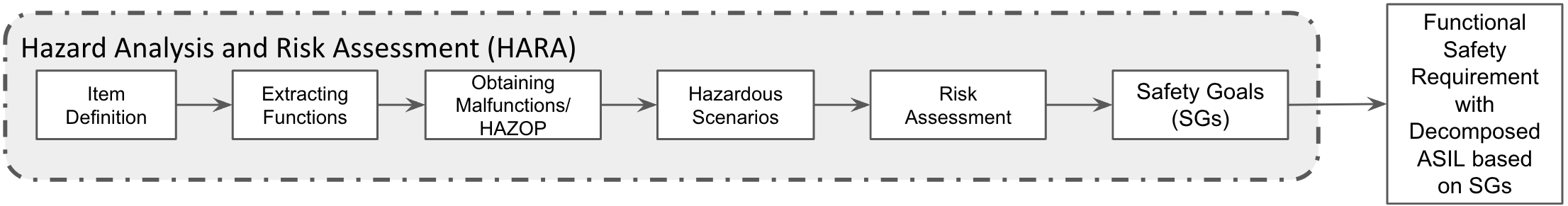}
    \caption{HARA safety verification follows this step by step procedure.}
    \label{fig:enter-label}
\end{figure*}

\noindent ASIL rating introduces a risk-based approach for categorizing risks into levels that dictate the necessary degree of risk reduction, from QM (Quality Management) to ASIL D, with ASIL D requiring the highest level of risk reduction \cite{kowalewski2013effort}. In this context, functions under ASIL A, B, or C, the standard mandates fewer requirements for development processes and safety mechanisms, while a QM rating suggests that standard quality measures expected in the automotive industry are sufficient. To systematically perform HARA, we separated HARA steps into the following as shown in Fig. \ref{fig:enter-label}.

\subsection{Item Definition}
\label{sec:2.1}
ISO 26262 calls for a detailed definition of the item under analysis, specifying its main functions and operating environment. This stage requires precise detail to ensure that the hazard analysis is practical and not overly complicated with excessive functions or sub-functions. The definition serves as a blueprint for understanding the item's interactions within its system and is crucial for identifying the roles and responsibilities that the item plays in its environment. In product development, specifically for ADS, this information could be extracted from product requirements and operational design domain (ODD) description. 

\subsection{Function Extraction}
\label{sec:2.2}
Extracting functions from item definitions in HARA involves identifying and defining the primary behaviors or actions of system components based on their descriptions (product requirements). This process requires a thorough understanding of item definitions, followed by the identification, documentation, and establishment of relationships between functions. By cross-referencing with requirements and engaging in an iterative, collaborative approach, engineers ensure that the derived functions accurately represent the intended system behavior, facilitating further analysis for identifying potential hazards and enhancing system safety and reliability.

\subsection{Malfunction Derivation}
\label{sec:2.3}
Malfunction derivation follows function extraction, where potential faults are identified using guide words inspired by the safety analysis. Guide words such as "no", "unintended", "early", "late", "more", "less", "inverted", and "intermittent" help developers contemplate all relevant faults in relation to the previously identified functions. However, not all fault types are relevant for every function; for example, the "more" fault type is not applicable to functions that produce a binary output. This step involves evaluating which fault types apply to each function and assessing the broader impact of each fault on the system.
\subsection{Hazards}
\label{sec:2.4}
In identifying hazards, each combination of faults and functions should be examined to identify all scenarios that could lead to potential hazards. This step focuses on understanding the vehicle-level behavior resulting from the item's malfunction, detailing the hazards and their possible impacts. Hazards, characterized by observable conditions or events noticeable by the driver, can be described verbally. This phase can also consider assumptions about driver actions to maintain control. Hazards are then linked to specific faults and situations, with stereotyped dependencies illustrating these relationships. Each relevant situation must be linked to at least one hazard, ensuring that all considered faults are associated with at least one hazard.

\subsection{Risk Assesment}
\label{sec:2.5}
Risk assessment in HARA is a critical process in ensuring the safety of autonomous driving systems. It involves evaluating potential hazards and determining the associated risks by analyzing three key factors: controllability, severity, and exposure.
\begin{itemize}
\item{\textbf{ Controllability (C)}} assesses the ability of the driver or system to prevent or mitigate the hazard. It is categorized into levels ranging from C0 to C3, with C0 indicating that the hazard can be easily controlled and C3 indicating that the hazard is difficult or impossible to control. For example, C0 means the driver or system can reliably avoid or mitigate the hazard, whereas C3 indicates a high likelihood of the hazard leading to a critical situation without effective intervention.

\item{\textbf{Severity (S)}} measures the potential impact of the hazard on human life, property, or the environment. Severity levels range from S0 to S3, with S0 indicating no injuries and S3 representing life-threatening or fatal injuries. For instance, S1 might involve minor injuries requiring medical attention, while S3 would involve severe injuries or fatalities.

\item{\textbf{Exposure (E)}} evaluates the frequency or likelihood of the hazard occurring during vehicle operation. Exposure levels range from E0 to E4, with E0 indicating an extremely unlikely event and E4 indicating a highly probable occurrence. For example, E2 represents a medium probability, meaning the hazardous situation could occur under certain conditions regularly encountered during normal driving.
\end{itemize}
By combining these factors, the safety engineer calculates an ASIL, which categorizes the risk and defines the necessary safety requirements. The ASIL ranges from A to D, with D representing the highest risk and most stringent safety measures. This systematic approach ensures that all potential hazards are addressed, and appropriate mitigations are implemented to maintain safety standards in autonomous driving systems. Table \ref{table:ASIL} shows the relation between C, E, S, and ASIL according to ISO 26262 \cite{ISO26262}.

\begin{table}[t!]
\centering
\renewcommand{\arraystretch}{1.4}
\setlength{\tabcolsep}{10pt}
\begin{tabular}{|c|c|c|c|c|}
\hline
\multirow{2}{*}{\textbf{Severity Class}} & \multirow{2}{*}{\textbf{Probability Class}} & \multicolumn{3}{c|}{\textbf{Controllability Class}} \\ \cline{3-5} 
                                         &                                             & \textbf{C1} & \textbf{C2} & \textbf{C3} \\ \hline
\multirow{4}{*}{S1}                      & E1                                          & QM          & QM          & QM          \\ \cline{2-5} 
                                         & E2                                          & QM          & QM          & QM          \\ \cline{2-5} 
                                         & E3                                          & QM          & QM          & A           \\ \cline{2-5} 
                                         & E4                                          & QM          & A           & B           \\ \hline
\multirow{4}{*}{S2}                      & E1                                          & QM          & QM          & QM          \\ \cline{2-5} 
                                         & E2                                          & QM          & QM          & A           \\ \cline{2-5} 
                                         & E3                                          & QM          & A           & B           \\ \cline{2-5} 
                                         & E4                                          & A           & B           & C           \\ \hline
\multirow{4}{*}{S3}                      & E1                                          & QM          & QM          & A           \\ \cline{2-5} 
                                         & E2                                          & QM          & A           & B           \\ \cline{2-5} 
                                         & E3                                          & A           & B           & C           \\ \cline{2-5} 
                                         & E4                                          & B           & C           & D           \\ \hline
\end{tabular}
\caption{ASIL determination table based on severity, probability, and controllability classes.}
\label{table:ASIL}
\end{table}

\subsection{Safety Goals}
\label{sec:2.6}
After conducting a risk assessment in the HARA process, safety goals (SG) are established to mitigate identified hazardous scenarios. These safety goals, defined as high-level requirements, aim to address hazards based on their ASIL rating, intended safe state, and fault-tolerant time interval (FTTI). They are derived from hazard analysis and ensure that each identified hazard with an ASIL rating has at least one corresponding safety goal. ISO 26262 mandates that the ASIL for a safety goal is set to the highest level among the hazards it addresses and specifies the fault tolerance time to refine the safety requirements further. These goals guide the development of technical safety requirements and design measures to address the root causes of hazards. By doing so, safety goals drive the system architecture and design, ensuring that potential risks are systematically mitigated. This approach enhances the overall safety of autonomous driving systems and ensures compliance with safety standards such as ISO 26262.

\section{HARA Integration with Generative AI}
\begin{figure*}[!h]
    \centering
 \includegraphics[width=0.9\textwidth]{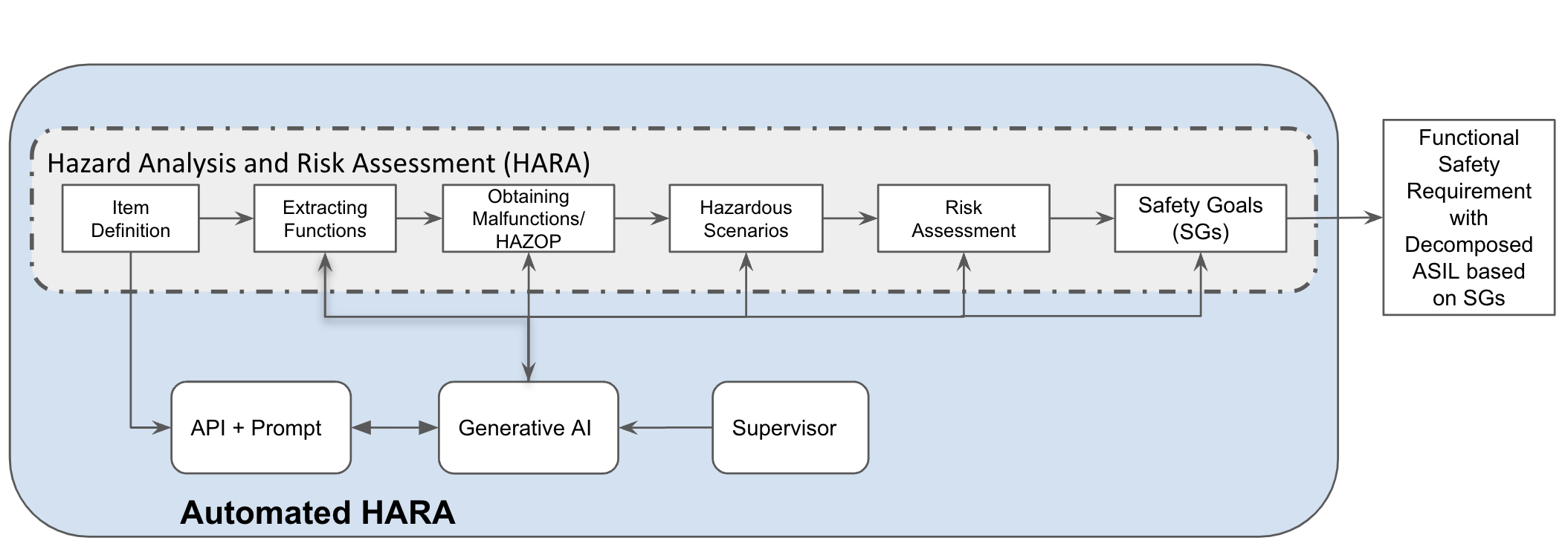}
    \caption{A schematic of the proposed automation architecture for HARA.}
    \label{fig:AutoHARA}
\end{figure*}

Performing HARA manually presents several significant challenges. The process is time-consuming, requiring extensive analysis and documentation to evaluate controllability, severity, and exposure for each identified hazard. Human error is another critical concern, as the complexity and volume of data can lead to oversight or misinterpretation of hazards and risks. Additionally, ensuring traceability throughout the HARA process is challenging, as it involves maintaining accurate records and justifications for every assessment and decision. Integrating generative AI into the HARA process can address these challenges by automating repetitive and data-intensive tasks, reducing the likelihood of human error, and enhancing traceability through systematic documentation. With the supervision of a safety engineer, generative AI can expedite the HARA process, ensuring more accurate and efficient risk assessments while maintaining high safety standards. Fig.\ref{fig:AutoHARA} shows the overall architecture of the generative AI integration to the HARA process.

The key components to ensure the effectiveness and reliability of automated HARA are:
\subsection{GUI}
 Firstly, a user-friendly Graphical User Interface (GUI) is essential to facilitate seamless interaction between the AI system and safety engineers. This GUI should support prompts that address each step of the HARA process, allowing the AI to generate initial assessments, recommendations, and documentation automatically.

\subsection{Supervision}
To ensure the integration is robust and adaptable, the system must enable safety engineers to refine AI-generated content at each step. This human-in-the-loop approach ensures that expert judgment and domain knowledge are applied, enhancing the accuracy and relevance of the assessments.

\subsection{Traceability} Additionally, maintaining a comprehensive record of changes and logs is crucial for traceability and accountability. The system should keep detailed records of all interactions, modifications, and decisions made throughout the HARA process. This audit trail is vital for compliance with safety standards such as ISO 26262 and for future reference during safety reviews and audits.

\subsection{Compatibility} Moreover, the AI integration should be designed to handle diverse data sources and formats, ensuring compatibility with existing tools and datasets used in safety engineering. By meeting these requirements, generative AI can significantly enhance the HARA process, making it more efficient, accurate, and traceable while maintaining rigorous safety standards.

\section{Benchmark Study on AEB}
To evaluate the efficacy of our proposed approach, we conducted a benchmark study on an ADS feature: Autonomous Emergency Braking (AEB). The primary objective was to compare the results of our AI-driven HARA approach with those derived from a traditional manual HARA process \cite{molloy2022safety}.
The benchmark study was designed to assess two main criteria:  coverage, and efficiency. In this study, two groups of safety engineers, each comprising four engineers, were assigned an identical task. One group conducted the Hazard Analysis and Risk Assessment (HARA) utilizing the Gen-AI framework introduced in this work, while the other group performed the analysis manually. In the following subsection, the system and the case study used in this work is explained.
\subsection{AEB System}
\begin{figure}[h!]
    \centering
 \includegraphics[width=0.4\textwidth]{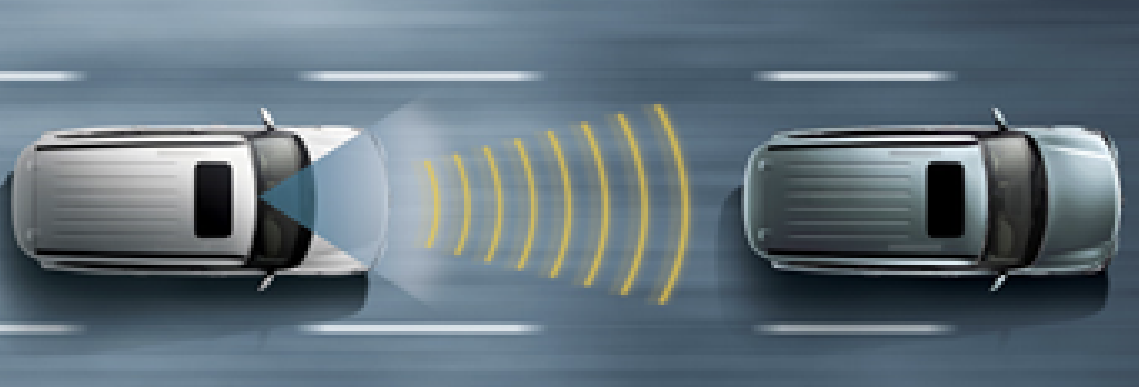}
    \caption{Collision avoidance using AEB system.}
    \label{fig:AEB_System}
\end{figure}

The AEB system is an advanced driver assistance feature designed to enhance vehicle safety by automatically detecting potential collisions and initiating braking to prevent or mitigate the impact. The system uses a combination of sensors, including radar, lidar, and cameras, to monitor the vehicle's surroundings. Upon detecting an imminent collision, the AEB system warns the driver through visual and auditory signals. If the driver fails to respond promptly, the system autonomously applies the brakes to avoid or lessen the severity of the collision. This functionality is crucial in reducing accident rates and improving overall road safety. Fig.\ref{fig:AEB_System} shows an example of the vehicle using a camera and radar to measure the distance and relative speed to be applied in AEB system.

In table \ref{table:product_requirements}, a list of the main product requirements for the AEB system is provided. Table \ref{table:ODD} provides the Operational design domain (ODD) that AEB should work for the vehicle. 
\begin{table}[h!]
\centering
\caption{Product Requirements for AEB System}
\begin{tabular}{c m{6cm}}
\hline\hline \\[-3mm]
\textbf{Requirement ID} & \textbf{Description} \\
\hline
PR1 & The system shall detect obstacles within a range of 150 meters. \\
\hline
PR2 & The system shall initiate braking within 300 milliseconds of predicting an imminent collision. \\
\hline
PR3 & The system shall apply sufficient braking force to bring the vehicle to a complete stop from a speed of 100 km/h within a distance of 40 meters. \\
\hline
PR4 & The system must provide a visual and auditory collision warning to the driver at least 2 seconds before initiating automatic braking. \\
\hline
PR5 & The system must adjust the braking profile based on the distance to the target and time to collision, ensuring a smooth and effective deceleration.  \\
\hline
\end{tabular}

\label{table:product_requirements}
\end{table}

\begin{table}[h!]
\centering
\caption{ODD for the AEB System}
\begin{tabular}{l m{6cm}}
\hline\hline \\[-3mm]
\textbf{ODD Parameter} & \textbf{Description} \\
\hline
Road Types & The system is designed for use on urban roads, highways, and rural roads. \\
\hline
Speed Range & The system operates effectively at speeds between 10 km/h and 150 km/h. \\
\hline
Weather Conditions & The system is operational in clear, rainy, snowy, and foggy conditions. It must also function in both day and night scenarios. \\
\hline
Traffic Density & The system is designed to operate in low to high traffic density environments. \\
\hline
Obstacles & The system must detect and respond to a variety of obstacles, including vehicles, pedestrians, cyclists, and static obstacles (e.g., barriers, parked cars). \\
\hline
Geographical Areas & The system must function in various geographical areas, including urban, suburban, and rural environments. \\
\hline
\end{tabular}
\label{table:ODD}
\end{table}

\subsection{HARA Using the Generative AI approach}
In this section, we conduct HARA using our proposed Generative AI approach for the AEB feature in autonomous driving systems. Based on the integration scheme shown in Fig.\ref{fig:AutoHARA}, generative AI is used to model and simulate diverse scenarios based on the given item definition and ODD. By leveraging advanced AI algorithms, this approach systematically identifies potential hazards, assesses associated risks, and suggests mitigation strategies. The steps of HARA will be explained in detail at each stage of this subsection, demonstrating how Generative AI can enhance hazard identification, risk assessment, and the development of effective mitigation strategies. This integration not only enhances the thoroughness of hazard identification but also improves the overall safety and reliability of autonomous driving systems.
\subsubsection{Function Extraction}
In this step, the generative AI model receives the product requirements and ODD presented in tables \ref{table:product_requirements} and \ref{table:ODD}, respectively. Based on the received information, the model extracts the functions related to each product's requirements. Table \ref{table:Functions} presents the extracted functions. The safety engineer supervises this process by reviewing and modifying the functions based on his/her expertise.

\begin{table}[h!]
\centering
\caption{Function Extraction Based on Product Requirements}
\begin{tabular}{cp{5cm}}
  \hline \hline \\ [-3mm]
  \textbf{Requirement ID} & \textbf{Function} \\
  \hline
  PR1 & Obstacle Detection \\
  \hline
 \multirow{2}{*}{PR2} & Collision Prediction \\
   & Braking \\
  \hline
  PR3 & Braking \\
  \hline
  PR4 & Collision Warning \\
  \hline
  PR5 & Braking\\
  \hline
\end{tabular}
\label{table:Functions}
\end{table}
\subsubsection{Obtaining Malfunction}
The malfunctions are obtained based on the driven functions in the previous steps. The combination of functions, HAZOP keywords \cite{dunjo2010hazard}, and generative capability of the AI model trained with safety-related data helps to obtain the potential malfunctions that could occur for the target system. Table \ref{table:Malfunctions} presents the malfunctions derived from the functions using the generative AI model. In this step, the safety engineer is responsible for reviewing and refining the malfunctions based on their expertise. This supervision at each step is required to avoid AI hallucination \cite{ji2023survey}.

\begin{table}[ht!]
\centering
\caption{Malfunction Extraction}
\begin{tabular}{m{3cm}m{6cm}}
\hline \hline \\[-3mm]
\textbf{Function} & \textbf{Malfunction} \\ \hline
\multirow{3}{3cm}{Obstacle Detection} & Obstacle not detected \\ 
& False Obstacle detected \\ 
& Delay on Obstacle Detection \\ \hline
\multirow{3}{3cm}{Collision Prediction} & Collision is not predicted \\ 
& False Collision is predicted \\ 
& Delay in collision prediction \\ \hline
\multirow{6}{3cm}{Braking} & Not braking \\ 
& Delay in braking \\ 
& Braking Stopped too soon \\ 
& Braking Stopped too late \\ 
& Too little braking \\ 
& Too much braking \\ 
& Braking too soon  \\ \hline
\multirow{3}{3cm}{Collision Warning}
& Not warning \\ 
& Too early warning \\ 
& Too late warning \\ 
& Stopped Warning too soon \\ 
& Provided Warning too long \\ 
& False warning \\ \hline

\end{tabular}
\label{table:Malfunctions}
\end{table}

\subsubsection{Hazardous Scenario}

The hazardous scenarios are obtained based on the malfunction, the ODD, and the generative capability of the AI model. In this step, the Generative AI model works as a brainstorming tool that helps the safety engineer come up with potential scenarios in which malfunction could lead to a hazard. Similar to the previous steps, the supervision of a safety engineer is required to prevent potential mistakes from the AI model. Table \ref{table:Hazards} shows the hazardous scenario driven from the malfunctions.

\begin{table*}[!h]
\footnotesize
\centering
\caption{Hazardous Scenario Derivation}
\begin{tabular}{m{25mm}m{10cm}}
\hline \hline \\ [-3mm]
\textbf{Malfunction} & \textbf{Hazard Description} \\ \hline
Obstacle not detected & AEB does not detect obstacle on Ego's path and front-end collision occurs at highway speed with the obstacle. \\ \hline
False Obstacle detected & AEB falsely detects an obstacle and Ego decelerates with maximum braking profile which may lead to rear-end collision with the following vehicle at highway speed. \\ \hline
Delay in Obstacle Detection & AEB delay in detection of obstacle may consume the time required for sufficient braking and lead to front-end collision with the obstacle. \\ \hline
Collision is not predicted & Failure to predict collisions leads to potential front-end collision with pedestrian/ road users\\ \hline
Delay in Collision Prediction &  Delay in Collision Prediction may lead to front-end collisions with pedestrians/ road users.\\ \hline
False prediction &  Incorrect collision prediction lead to unnecessary braking and may lead to rear-end collision with vehicles following Ego.\\ \hline
Not braking & AEB failure to apply braking when Ego is on the collision path may lead to front-end collision with pedestrian/ road users.\\ \hline
Delay in Braking & Delayed response in applying brakes may lead to potential front-end collision with pedestrian/road users \\ \hline
Braking stopped too soon & Stopping braking too soon when Ego is still in collision path resulted in inadequate deceleration and may lead to front-end collision with pedestrian /road-users \\ \hline
Too little braking&  Sluggish response in applying brakes leading to potential collision with road users \\ \hline
Too much braking & Braking more than the required deceleration to avoid collision may destabilize the vehicle and lead to side collisions with adjacent road users. \\ \hline
Braking too soon & Braking too soon when the collision threat is not imminent may lead to rear-end collision with the following vehicle \\ \hline
Not warning &  Front-end collision due to failure to warn the driver for potential collision when AEB cannot avoid the collision.\\ \hline
Too early warning & Giving warning too early may lead to loss of trust of the driver and not taking back control when the real collision threat is ahead. \\ \hline
Too late warning & Too late warning may consume the required time for the driver to take back control and perform evasive maneuvers. \\ \hline
Stopped warning too soon & Stopping the warning too soon when there is a collision threat may lead to front-end collision by not performing evasive maneuvers by the driver. \\ \hline
Provided warning too long & Driver becomes complacent due to excessive warning time and disregards future warnings which may lead to collision when there is an imminent collision threat. \\ \hline
False Warning & Driver experiences false positive warnings leading to driver complacency or disregard for future legitimate warnings which may lead to collision when driver evasive maneuver is required to avoid collision \\ \hline
\end{tabular}
\label{table:Hazards}
\end{table*}

\subsubsection{Risk Assesment}
In this step, the Generative AI model based on the hazardous scenario, ODD, traffic information, historical collision data, and the potential collision comes up with the Controlability (C), Exposure (E), and Severity (S). Then, based on the information given in Table \ref{table:ASIL}, the ASIL allocated to each hazard will be determined.

\begin{table*}[!h]
\footnotesize
\centering
\caption{Safety Goals}
\begin{tabular}{m{0.3cm}m{6.3cm}m{4cm}m{0.5cm}}
\hline \hline \\ [-3mm]
\textbf{No.} & \textbf{Hazardous Scenario} & \textbf{Safety Goal} & \textbf{ASIL} \\ 
\hline
1 & AEB does not detect obstacles on Ego's path and front-end collision occurs at highway speed with the obstacle. & SG 1: The AEB system shall detect all obstacles on the Ego's path. & D \\ 
\hline
2 & AEB falsely detects an obstacle and Ego decelerates with maximum braking profile which may lead to rear-end collision with the following vehicle at highway speed. & SG 2: The AEB system shall avoid false detections to prevent unintended braking. & C \\ 
\hline
3 & AEB delay in detection of obstacle may consume the time required for sufficient braking and lead to front-end collision with the obstacle. & SG 3: The AEB system shall ensure timely detection of obstacles to allow sufficient braking time. & D \\ 
\hline
4 & Failure to predict collisions leads to potential front-end collision with pedestrian/ road users. & SG 4: The system shall ensure prompt collision prediction. & D \\ 
\hline
5 & Delay in Collision Prediction may lead to front-end collisions with pedestrians/ road users. & SG 4: The system shall ensure prompt collision prediction. & D \\ 
\hline
6 & Incorrect collision prediction leads to unnecessary braking and may lead to rear-end collision with vehicles following Ego. & SG 5: The AEB system shall avoid false collision prediction to prevent unintended braking. & C \\ 
\hline
7 & AEB failure to apply braking when Ego is on the collision path may lead to front-end collision with pedestrian/ road users. & SG 6: The AEB system shall apply braking when the Ego vehicle is on a collision path. & D \\ 
\hline
8 & Delayed response in applying brakes may lead to potential front-end collision with pedestrian/road users. & SG 7: The AEB system shall ensure a timely braking response. & D \\ 
\hline
9 & Stopping braking too soon when Ego is still in collision path resulted in inadequate deceleration and may lead to front-end collision with pedestrian/ road-users. & SG 7: The AEB system shall ensure a timely braking response. & D \\ 
\hline
10 & Sluggish response in applying brakes leading to potential collision with road users. & SG 7: The AEB system shall ensure a timely braking response. & D \\ 
\hline
11 & Braking more than the required deceleration to avoid collision may destabilize the vehicle and lead to side collisions with adjacent road users. & SG 8: The AEB system shall avoid vehicle destabilization during the braking. & C \\ 
\hline
12 & Braking too soon when the collision threat is not imminent may lead to rear-end collision with the following vehicle. & SG 7: The AEB system shall ensure a timely braking response. & C \\ 
\hline
13 & Front-end collision due to failure to warn the driver for potential collision when AEB cannot avoid the collision. & SG 9: The AEB system shall warn the driver at least 2 seconds before engaging the braking. & D \\ 
\hline
14 & Giving a warning too early may lead to loss of trust of the driver and not taking back control when the real collision threat is ahead. & SG 10: The system shall optimize the timing of warnings to maintain driver trust and ensure appropriate control handover. & QM \\ 
\hline
15 & Too late warning may consume the required time for the driver to take back control and perform evasive maneuvers. & SG 9: The AEB system shall warn the driver at least 2 seconds before engaging the braking.& D \\ 
\hline
16 & Stopping the warning too soon when there is a collision threat may lead to front-end collision by not performing evasive maneuvers by the driver. & SG 11: The system shall ensure warnings persist until the collision threat is resolved. & A \\ 
\hline
17 & Driver becomes complacent due to excessive warning time and disregards future warnings which may lead to collision when there is an imminent collision threat. & SG 12: The system shall avoid false warnings to maintain driver trust and ensure responsiveness to legitimate warnings.& QM \\ 
\hline
18 & Driver experiences false positive warnings leading to driver complacency or disregard for future legitimate warnings which may lead to collision when driver evasive maneuver is required to avoid collision. & SG 12: The system shall avoid false warnings to maintain driver trust and ensure responsiveness to legitimate warnings. & QM \\ 
\hline
\end{tabular}
\label{table:SafetyGoals}
\end{table*}

\subsubsection{Safety Goals}
In this step, the safety goals required to prevent the driven hazardous scenarios are generated. These safety goals will inherit the ASIL for the hazards and will be linked to them. Throughout this step, each hazard should be linked to at least one safety goal.
Table \ref{table:SafetyGoals} presents the safety goals driven using the generative AI model based on the hazardous scenarios obtained in the previous steps. 
Overall, 12 safety goals are driven of which 10 safety goals have ASIL and need further attention to decompose them to the safety requirements using the conventional methods in safety engineering.\\
It should be noted that a safety engineer supervised this step and modified the safety goals when necessary. The advantages of using the AI model are reducing the required time to obtain the safety goals and helping to keep the traceability between different steps.

\section{Results and Discussion}
The results show significant improvements in both time efficiency and analysis coverage with our AI-driven approach.

\subsection{Time Efficiency}

The integration of generative AI into the HARA process substantially reduced the time required to perform safety analysis. Compared to the traditional manual approach, which is both time-consuming and labor-intensive, the AI-driven HARA demonstrated an impressive reduction in analysis time by up to 80\%. This improvement highlights the developed approach's capability to swiftly process large datasets and simulate a multitude of scenarios, which traditionally would require extensive human effort and time. This is illustrated in Table \ref{tab:coverage_improvement} and Figure \ref{fig:hara_comparison}. The significant time savings achieved through the AI-driven method demonstrate its potential to streamline safety analysis workflows, making the process more efficient without compromising on thoroughness.

\begin{table}[!h]
\centering
\caption{Coverage and Time Improvement Comparison}
\label{tab:coverage_improvement}
\begin{tabular}{lcc}
\hline \hline \\[-3mm]
\textbf{Method} & \textbf{Coverage (\%)}& \textbf{Time Required (hours)} \\ \hline
Traditional Manual HARA & 80 & 100\\ 
AI-driven HARA & 100 & 20\\ \hline
\end{tabular}
\end{table}

\subsection{Coverage Improvement}

\begin{figure}[ht!]
    \centering
    \includegraphics[width=0.45\textwidth]{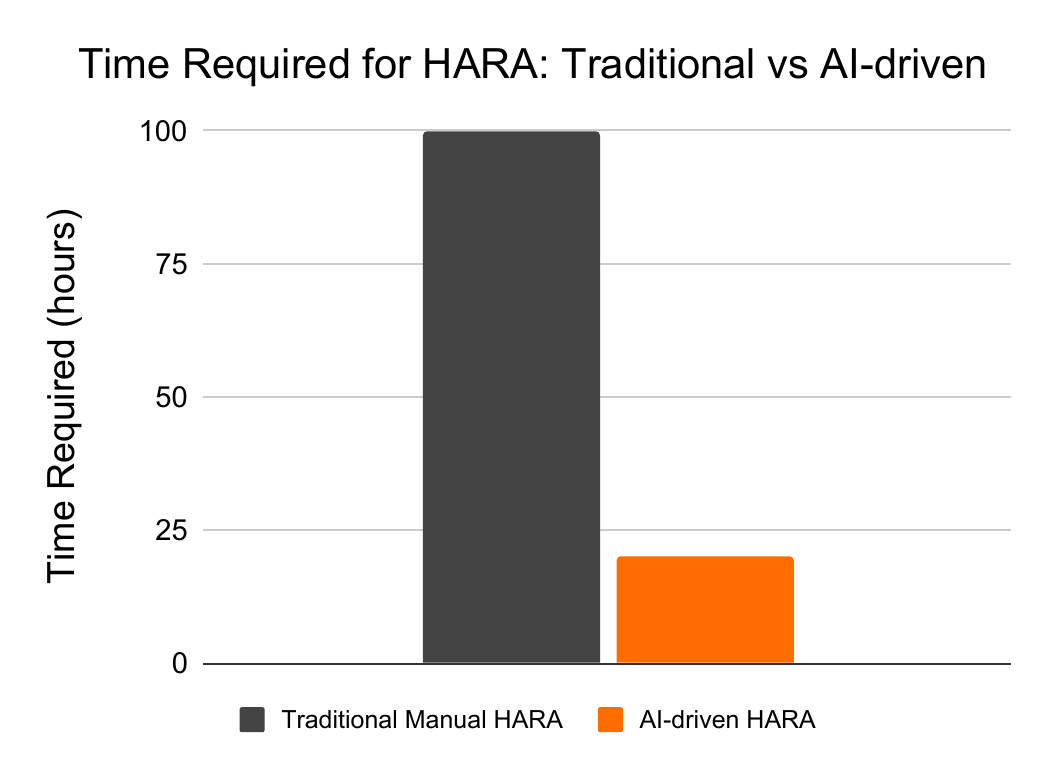}
    \includegraphics[width=0.45\textwidth]{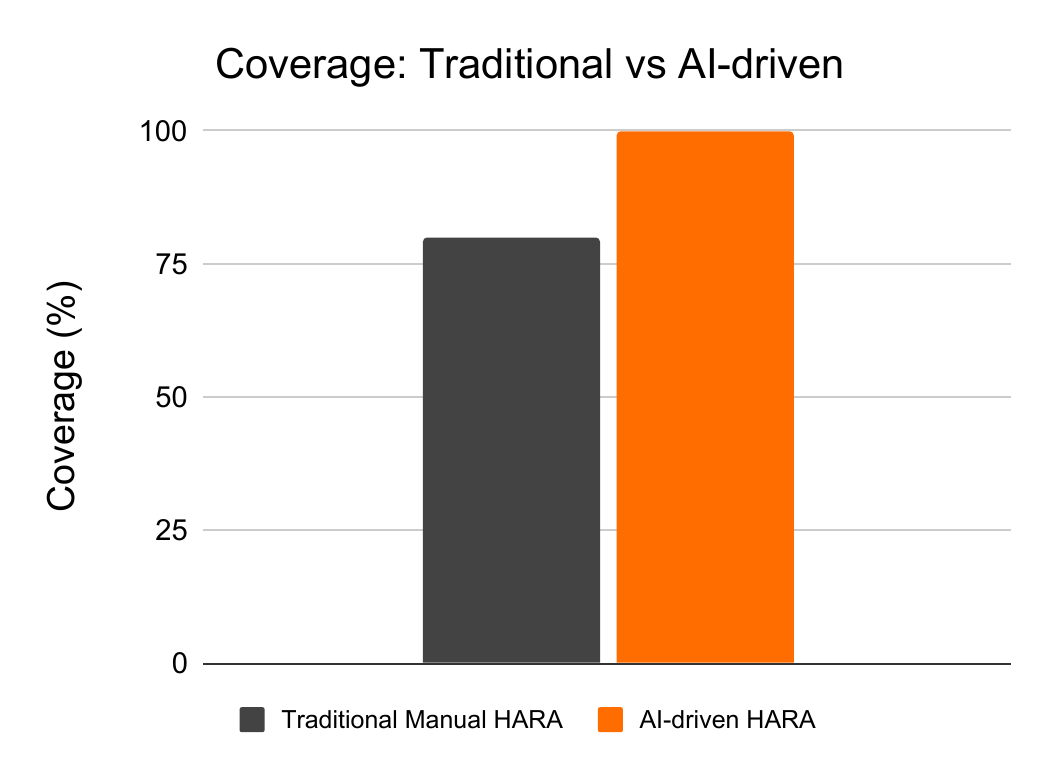}
    \caption{Comparison of Time Required and Coverage: Traditional vs AI-driven HARA}
    \label{fig:hara_comparison}
\end{figure}

Our AI-driven HARA approach has delivered notable enhancements in the coverage of hazard identification. On average, the developed AI-enhanced process identified 20\% more hazards than the conventional manual approach conducted by safety engineers. This increase in coverage is decisive, as it ensures a more comprehensive identification of potential risks, including those that might be overlooked in traditional analyses. This is detailed in Table \ref{tab:coverage_improvement} and Figure \ref{fig:hara_comparison}. By systematically evaluating a wide range of scenarios, including rare and complex ones, our tool broadened the horizon of hazard perception and provided innovative insights into potential risk factors that could impact autonomous driving systems.


\subsection{Traceability and Auditing}

In addition to improving time efficiency and coverage, the platform significantly enhances traceability and auditing capabilities within the HARA process. The platform maintains detailed logs of all changes, decisions, and assessments made throughout the analysis, hence ensuring that the safety analysis is thorough, fully documented, and easily auditable. The ability to track and review every step of the HARA process facilitates compliance with safety standards such as ISO 26262, providing a clear audit trail that supports accountability and continuous improvement in safety engineering practices.

\section{Conclusion}

Generative AI transforms HARA by streamlining workflows, mitigating oversight, and broadening hazard perception. Integrating generative AI in the early stages of safety system development ensures proactive identification and mitigation of latent system vulnerabilities. This fosters a robust safety culture aligned with the complexities of autonomous driving technologies.

The benchmark study on the AEB system demonstrates that our AI-driven HARA tool significantly enhances the efficiency and effectiveness of hazard analysis. Compared to the traditional manual process, the proposed AI-driven approach reduced the required time for performing HARA by up to 80\% and improved the coverage by 20\%. Additionally, our platform maintains comprehensive traceability and change logs, facilitating easier auditing and review processes.

These improvements highlight the potential of generative AI in advancing safety analysis processes. By leveraging generative AI, safety engineers can achieve more thorough and timely hazard assessments, ultimately contributing to the development of safer and more reliable autonomous driving systems.

\bibliographystyle{IEEEtran}
\bibliography{IEEEabrv,Bibliography}

\end{document}